# Tunable supramolecular polymerization from protein charge heterogeneity and architecture

Nynke Marije Hettema[†1], Meng Shen[†2*], Frank van Opstal[1], Emmett Lim[2], Liedewij Laan[1*]

[†] These authors contributed equally to this work
[1] Bionanoscience Department, Delft University of Technology, Delft 2629HZ, The Netherlands
[2] Department of Physics, California State University Fullerton, Fullerton, CA 92831, USA
*Corresponding Authors: Liedewij Laan and Meng Shen

**Email:** L.laan@tudelft.nl and meshen@Fullerton.edu

**Abstract**

Multidomain proteins with flexible unstructured sequence regions are abundant in cellular signaling processes. This protein architecture enables self-assembly into supramolecular structures, but how structured interaction domains and overall protein architecture jointly regulate the assembly size, structure and kinetics remains unclear. Here we use the budding yeast protein Bem1 as a model multidomain system to show that supramolecular polymerization can be tuned by charge heterogeneity and protein architecture. We experimentally demonstrate that Bem1's isolated PB1 domain forms extended filaments, whereas full-length Bem1 forms substantially shorter assemblies, indicating that the PB1 domain drives assembly while the remaining protein architecture tunes filament length. To understand these observations, we develop minimal coarse-grained models approximating the PB1 as a polar 5-bead domain and the full-length Bem1 as a 6-bead model with an additional bead representing the remainder of Bem1. The weight distribution of supramolecular filaments assembled by the 5-bead model quantitatively follows reversible Flory-like polymerization theory, which is tunable within a narrow charge polarity regime. In contrast, the 6-bead model shifts chain-length distributions towards shorter polymers despite retaining the same driving domain. We show that this deviation arises from steric and geometric constraints imposed by the appended unstructured regions, where the rotational flexibility between the charge-polar structured domain and the unstructured region emerges as the key physical parameter governing self-limited self-assembly. Together, our results establish charge polarity, protein architecture, and conformational flexibility as programmable control knobs for supramolecular polymerization and suggest a general framework for understanding how multidomain proteins assemble into tunable biomolecular structures.

**Significance Statement**

Protein self-assembly of multidomain proteins underlies many cellular signaling processes. We identify two tunable physical parameters that drive and tune protein polymerization: polar charge heterogeneity in folded domains and rotational flexibility in overall protein architecture.
Polar charge heterogeneity can drive polymerization where the charge carried in opposite poles of a structured domain can tune polymer length and flexibility. These characteristics can be further tuned by the overall protein structure and rotational flexibility. Using experimental measurement of self-assembly by yeast polarity protein Bem1 and its polar folded domain PB1 combined with coarse-grained molecular dynamics simulations, we show the relevance of these two tuning parameters and build a general framework for tunable self-assembly in multidomain proteins that are ubiquitous in cellular signaling.

## Introduction

Cellular signaling, essential for cell growth, division, motility or to respond to internal and external cues (1), often emerges from the local enrichment of many interacting molecules at a specific location (1–3), resulting in a tightly coordinated spatiotemporal control during the cell cycle(1, 3). The proteins involved in these processes often have a combination of several folded functional domains and flexible unstructured sequence regions (4–8), providing architectures capable of multimerization or self-assembly (9), conformational regulation (10–12) or context-dependent interactions necessary for complex functionalities (13, 14).

Previous studies have shown that structured domains can drive higher-order assembly through interaction motifs, leading to polymeric or chain-like macromolecular structures (15–20). Conversely, flexible disordered or unstructured regions can modulate association and dissociation kinetics (10, 21, 22), alter steric accessibility (23), and suppress ordered aggregation (24–26). Meanwhile, electrostatics plays an important role in biomolecular self-assembly (27, 28), which can be controlled by charge heterogeneity of proteins and solvent environment (29–33), and potentially leading to dynamic self-assembly of supramolecules that are highly responsive to external stimuli (34–36). Despite these interesting observations, much less is understood about how structured domains and flexible unstructured regions cooperate within a single multidomain protein to form supramolecular polymers.

In particular, two questions remain to be addressed. First, how does heterogeneous charge distribution within structured domains influence the structure, size and dynamics of supramolecular self-assembly? Second, how do flexible unstructured regions regulate continued self-assembly of a multidomain protein?

To address these questions, we focused on cell polarity in *Saccharomyces cerevisiae* as a model signaling process. We used a literature-curated set of 51 cell polarity proteins (37) to compare them with the full yeast proteome. Indeed, polarity proteins are more enriched in multidomain architectures, while also more frequently containing substantial unstructured regions (Figure 1A), consistent with the structural complexity of signaling proteins, described above. Among these proteins, Bem1 combines multiple folded domains, including a highly charge-polar domain, and substantial flexible unstructured regions (38–40). Specifically, Bem1 contains two SH3 domains and a PX domain positioned towards the center portion of the protein, connected through an unstructured linker to a C-terminal PB1 domain with highly heterogeneous and strongly polar charge distribution regions (Figure 1B) (38). These features make Bem1 an ideal candidate for elucidating how domain charge heterogeneity and global protein architecture jointly control supramolecular self-assembly. Therefore, we choose Bem1 as a biologically motivated model system.

Recent sticker–spacer frameworks have provided powerful descriptions of how sequence patterning in disordered proteins controls phase separation and condensate properties (41–43). Whereas sticker-spacer model pioneered sequence-based descriptions of biomolecular self-assembly, we design a minimal model framework to focus on emergent physical mechanisms arising from electrostatic interactions and protein architecture at a more coarse-grained physical level.

Here, we combine *in vitro* experiments with minimal coarse-grained simulations. We carried out size-exclusion chromatography with multi-angle light scattering (SEC-MALS), western blot, and negative stain transmission electron micrographs (nsEM) on both full-length Bem1 and PB1-only to show self-assembled supramolecular polymer formation. Subsequently we performed minimal coarse-grained molecular dynamics (MD) simulations to show how structured domain charge heterogeneity and the flexible unstructured sequence regions of a full-length protein together can tune the observed supramolecular assemblies.

## Results

### Bem1 self-assembles into a variety of multimers in solution

To study the behavior of multidomain protein Bem1 at high local protein density, as occurs at the cell polarity spot (**Figure 1A**), we carried out SEC-MALS on purified full length Bem1 (**Figure 1B, Figure S1**) (His-Bem1-Flag, from now on referred to as Bem1). The concentration profile shows multiple peaks throughout the elution range (**Figure 1C**), indicating supramolecular protein complexes of a wide size range. Multi-angle light scattering measurements show that the molecular mass in the peak fractions ranges from monomers(70kDa) to 18-mers(1260kDa). Western blot analysis of separate SEC elution fractions confirms that all peaks in the concentration profile contain Bem1 monomers (70kDa) (**Figure 1C**), indicating that the supramolecular complexes are made of protein monomers. Based on the concentration profile (dRI), Bem1 shows biases towards its monomeric and trimeric state. Dimers are present in the shoulder of the trimer peak. Higher order multimers (tetramer to 20-mer) cannot be distinguished in separate peaks and together make up the remaining protein mass. The statistical average of the calculated mass distribution of 6 SEC-MALS traces shows a similar trend where around 40% of Bem1 is in the monomeric fraction after which multiple peaks of higher order multimers are present of which the mass distribution declines with exponential decay with a long tail (**Figure 1D**). Large variation is present between the individual traces, especially pronounced in the trimer to septamer range. This large variation between experiments further supports the hypothesis that transient protein interactions drive Bem1 multimer formation.

To investigate what these multimers look like we turned to negative stain EM (nsEM), which revealed that Bem1 forms a range of different amorphous shapes. Many of the larger structures show an elongated trend (**Figure 1E**), with a smaller set of structures showing pronounced long structures of over 200 nm. The long structures show strong bends and are irregular in width (**Figure 1E**). The elongated morphology suggests anisotropic interactions between Bem1, plausibly associated with the heterogeneous charge distribution of the protein. The broad range of multimer sizes observed in nsEM extends the multimeric population detection in SEC-MALS to longer-chain-like assemblies, confirming larger multimers can form in the absence of flow-induced shearing during SEC-MALS. This polydispersity is consistent with reversible polymerization driven by physical interactions rather than a single stoichiometric complex. In addition, the diversity of elongated shapes suggests the assembled supramolecular can adopt various conformations, implicating flexible unstructured regions of Bem1 in modulating the assembly structures. Together, the SEC-MALS and nsEM measurements indicate that Bem1 monomers spontaneously assemble into dynamic multimers or supramolecules under both flow and quiescent conditions.

### PB1 protein domain drives Bem1 multimerization

To understand the mechanisms underlying the multimerization of Bem1, we first analyzed Bem1's charge distribution based on its AlphaFold-predicted structure by electrostatic calculations performed with the Adaptive Poisson-Boltzmann Solver (APBS) in Pymol (44). Although the overall charge of +6e in Bem1 does not seem significant considering overall 140 charged residues (-67e and +73e) (45), the charge distribution in structured PB1 domain is spatially heterogeneous, with one region of the PB1 domain enriched by negatively charged residues (ASP+GLU) and the opposite region enriched with positively charged residues (ARG+LYS), leading to a strongly polar surface charge distribution in PB1 (**Figure 1B, iv**). The orientation of this charge polarity is consistent with type II/type B PB1 domains (39, 46). Moreover, the C-terminal PB1 domain is preceded by a flexible unstructured region in Bem1 (38), allowing accessibility for (self-)interactions in full-length Bem1.

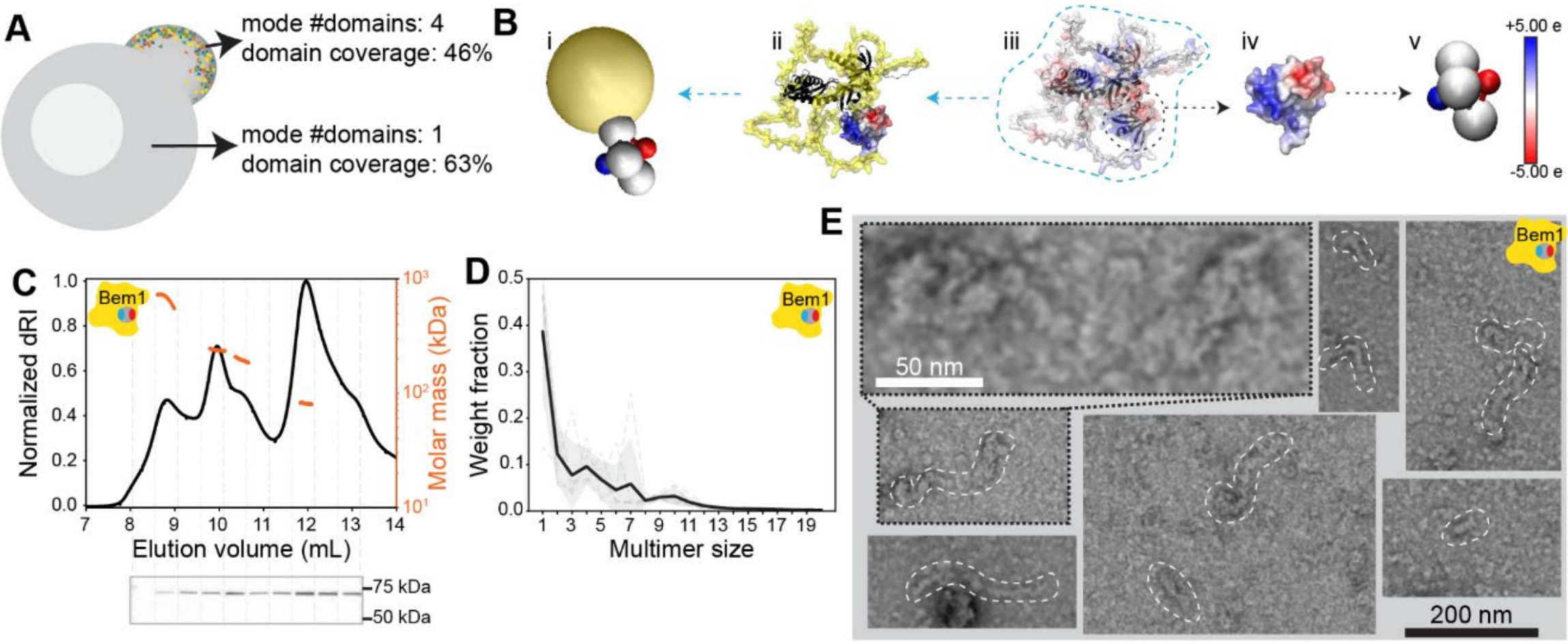

**Figure 1. Bem1 self-assembles into a variety of multimers in solution. A)** Cartoon representation of a dense region of signal establishment in a polarized yeast cell with a comparison of two protein properties between the full yeast proteome (n=6766) and the polarity proteins set (n=51): mode number of folded domains and domain coverage of the protein sequence. **B)** (i) 6-bead model representation of full-length Bem1. (ii) Charged surface visualization of PB1 surrounded by the remaining structure of Bem1, unstructured regions highlighted in yellow. (iii) Protein structure of Bem1 as predicted by AlphaFold, visualized as cartoon with charged surface. (iv) Charged surface visualization of isolated PB1 domain. (iii) 5-bead model representation of PB1. (v) Charged surface visualization of PB1 surrounded by the remaining structure of Bem1, unstructured regions highlighted in yellow. **C)** SEC elution profile (solid black line) and MALS trace (orange lines) for selected peaks (TOP) and Western blot analysis of SEC-MALS elution fractions (BOTTOM) for purified Bem1 in solution. **D)** Weight distribution of Bem1 multimers. Mean of n=6 SEC-MALS traces (solid black line), individual traces (dotted grey lines), and 95% confidence interval (CI) using the t-distribution (shaded area). **E)** negative-stain transmission electron micrograph of Bem1 multimers in solution. White dashed lines indicate observed elongated structures, scalebar is 200nm. Zoomed-in panel of Bem1 structure indicated by black dashed line, scalebar 50 nm.

These structural features inspired us to conceptually split the Bem1 protein into two different parts: i) the charge-polar structured PB1 domain, and ii) the remaining unstructured and structured sequence that is connected to the PB1 domain by a flexible unstructured region (38). To isolate their respective roles, we purified the isolated PB1 domain (His-Bem1$_{478\text{-}551}$-Flag, from now on referred to as PB1), (**Figure 1B iv**) and a Bem1 mutant without the PB1 domain (His-Bem1$_{1\text{-}464}$-Flag, from now on referred to as Bem1ΔPB1) (**Figure S1**). We then compared the self-assembly properties of PB1, Bem1ΔPB1 and full-length Bem1. SEC-MALS analysis of Bem1ΔPB1 reveals a dominant main peak containing ~97% of total protein mass and a small secondary peak comprising the remaining 3% mass fraction (**Figure 2A**). Western blot analysis and MALS mass analysis identify the main peak as predominantly Bem1ΔPB1 monomers (60 kDa), while the minor peak corresponds to Bem1ΔPB1 dimers. Together, these results show that the Bem1ΔPB1 strongly favors the monomeric state, indicating the non-PB1 parts of Bem1 alone do not drive substantial multimerization and supporting the hypothesis that the PB1 domain is the dominant assembly-driving domain.

We also attempted SEC-MALS analysis of the purified PB1 domain, however, self-assembly into large clusters prevented measurements due to interference of the chromatographic flow, indirectly indicating stronger self-assembly by PB1 alone than full-length Bem1. Consistent with this implication, nsEM images reveal more abundant, substantially longer and thinner filaments formed by purified PB1 domains than full-length Bem1 (**Figure 2B**). These filaments have a characteristic width of 3nm, markedly thinner than elongated assemblies formed by full-length Bem1, and consistent with the approximate hydrodynamic size of a single PB1 domain (46). Their contour lengths span a broad distribution up to several micrometers with the majority of filaments in the range up to 250 nm, corresponding to approximately 2-83 PB1 monomers per chain, suggesting head-to-tail PB1 polymerization (16). Given sample handling is expected to break up larger filaments, even longer filaments are expected in solution.

Head-to-tail polymerization has been previously reported for other PB1 variants for a combined type I/II structure containing both the characteristic lysine residue in β-sheet 1 and a conserved acidic OPCA motif (16, 34, 46, 47). Although the Bem1 PB1 domain is classified as type II and lacks the canonical acidic OPCA motif (39), other charged surface patches may provide interaction sites, analogous to the OPCA motif, to drive polymerization in a similar fashion as the type I/II structure (**Figure 1B**).

The experimental results also provide insights into the physical nature of PB1 self-assembly. First, because PB1 lacks Cys residues in the PB1 structure (48), disulfide-bond-mediated crosslinking is unlikely. Second, the broad polydisperse filament weight distribution (**Figure 2C**) is characteristic of reversible supramolecular polymerization rather than formation of a single stoichiometric complex. Third, varying ionic strength during self-assembly showed that decreasing salt concentration twofold increased both the abundance and length of PB1 assemblies and promoted extended filamentous networks including much longer chain lengths (**Figure S2**). Together, these observations support a noncovalent assembly mechanism with an important electrostatic contribution.

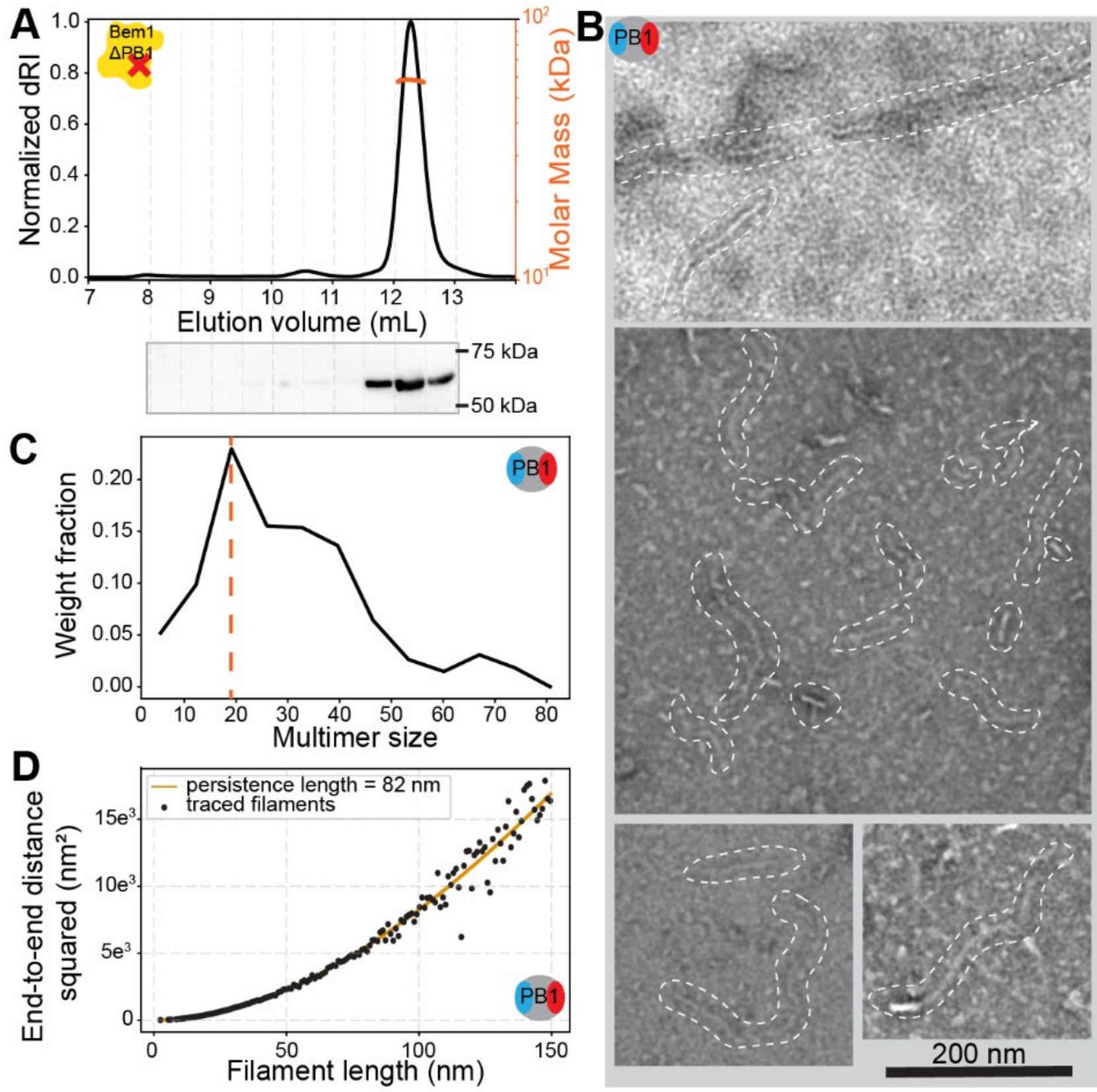

**Figure 2. The PB1 protein domain is a driver for Bem1 multimerization through polymerization. A)** Size exclusion profile (solid line) and MALS trace(dotted line) for selected peaks (TOP) and Western blot analysis of SEC-MALS elution fractions (BOTTOM) for $Bem1_{1\text{-}464}$ (Bem1ΔPB1) in solution. **B)** negative-stain transmission electron micrograph of $Bem1_{478\text{-}551}$ (PB1 domain). Scalebar is 200nm, dashed lines indicate observed filaments. **C)** Weight distribution of PB1 filaments, measured from n=250 PB1 structures from nsEM. Mean value (19-mer) displayed by the orange dashed line. **D)** End-to-end distance squared vs traced filament length fitted of n=150 PB1 structures from nsEM micrographs with the expected curve for persistence length 82nm.

Finally, joint analysis of the filament contour length and the end-to-end distance yielded a persistence length of 82±7 nm ($R^2$=0.9620) (**Figure 2D**), which roughly corresponds to 30 single PB1 domains, indicating that PB1-assembled filaments possess relatively higher bending rigidity then the full-length Bem1 supramolecules, but a relatively lower bending rigidity than typical cellular filaments like cytoskeletal components or flagella (49).

The comparison between thin and long filaments assembled by PB1 and the thicker, shorter and less abundant filaments formed by Bem1 motivated us to hypothesize that the self-assembly of Bem1 supramolecular conformations is driven by charge heterogeneity in structured PB1 domains, while the remaining complex of structured and unstructured regions inhibit continued polymer assembly. To test this hypothesis and its transferability to other multidomain protein systems, we resort to minimal coarse-grained molecular dynamics simulations.

**Experimentally-inspired minimal coarse-grained models: 5- bead model and 6- bead model**

To test the generality of the experimental motivated hypothesis of different roles of structured domain, flexible unstructured regions and charge heterogeneity in multidomain protein self-assembly, we construct two minimal coarse-grained models (**Figure 1B i and v**) as follows. In the first model, the 5-bead model representing a structured charge-polar domain (PB1-like), is represented by 5 beads arranged at the vertices of a triangular bipyramid (**Figure 1B v**). Two smaller beads located at opposing vertices carry positive and negative charges, respectively, mimicking the polar charge distribution of the PB1 domain. The remaining three larger, neutral beads form the central triangular plane and represent the excluded volume of the structured PB1-like domain, preventing unphysical interpenetration of assembly units. The second model extends the 5-bead model by connecting it to an additional, larger neutral bead, yielding the 6-bead model (**Figure 1B i**). This added bead approximately represents the remainder of the Bem1-like protein, lumping together the structured and unstructured regions. Given charge heterogeneity is substantially weaker in the remaining regions of full-length Bem1, the 6-bead model neglects charge distribution outside PB1 domain for simplicity. Unless otherwise specified, the 5-bead domain is allowed to freely rotate with respect to the additional larger bead, mimicking the flexible connection between PB1 and the rest of Bem1 by a flexible unstructured region. This extension allows us to isolate the steric hindrance effect of non-PB1-like regions without introducing additional electrostatic complexity.

Together, these minimal 5-bead and 6-bead models enable large-scale simulations of protein self-assembly that generate statistically representative data for extracting relevant parameters for phenomenological polymerization theories to test mechanistic hypotheses while retaining the essential physical ingredients required to reproduce experimental trends.

Three independent control parameters allow us to tune the structure and kinetics of self-assembly:

i) $q$: the charges carried by the two opposing poles;

ii) $d_{np}$: The distance between the two charged beads (poles);

iii) $d_{others}$: the distance between the triangular bipyramid and the larger neutral bead in the 6-bead model;

iv) $k_{angle}$: rotational stiffness between the triangular bipyramid and the larger neutral bead.

These parameters independently control the interaction strength and dipolar geometry of PB1-like domains and steric hindrance from non-PB1-like regions, respectively.

**Pole charges of major structured domain tune the supramolecular polymerization kinetics and thermodynamics**

We begin by examining the self-assembly of the 5-bead model. In this minimal description, the assembly behavior of a charge polar structured domain is governed by two effective parameters, the pole charge $\pm q$, and the pole-to-pole distance $d_{np}$.

We hypothesize that the pole charge $\pm q$ controls the strength of inter-monomer attraction and thus governs both the kinetics and thermodynamics of the self-assembly process. To isolate the effect of pole charges on self-assembly, we systematically vary the pole charge from 3 $e$ to 13 $e$ while fixing $d_{np} = 2.24$ nm. Snapshots in **Figure 3A, Figure S4B to D and Supplemental Videos S1 to S3** show that the 5-bead models with these parameter sets indeed self-assemble to chain-like supramolecule structures rather than forming 3D structures.

The self-assembly of supramolecules is further quantified by the weight distribution of self-assembled supramolecules, $w_N$:

$$w_N = \frac{n_N N}{\sum_{N'} n_{N'} N'} \tag{1}$$

where $N$ is the number of monomers in each polymer and $n_N$ the number of N-mers. Zoomed-in snapshots taken after 20 $\mu s$ simulations (**Figure 3C to E**) show the formation of filaments with weight distribution of supramolecules progressively shifting towards longer chain length for pole charges $q = \pm 3.8\ e, \pm 4.0\ e, \pm 5.0\ e$, respectively (**Figure 3B**). Interestingly, the weight distribution of supramolecules for $q = \pm 13.0\ e$ almost coincides with that for $q = \pm 5.0\ e$ (**Figure S5A**), indicating that the tunability of pole charge on self-assembly saturates for $|q| \geq 5.0\ e$.

Although 20 $\mu s$ is relatively long for molecular dynamics simulations, it remains orders of magnitude shorter than the accessible time resolution in experiments. To bridge this gap, we assess simulation results with polymerization theories. Within the Flory theory, polymerization is fully characterized by the single parameter, $p$, provided that $p$ is independent of the bond's position along the chain (50). Under this assumption, the weight fraction of N-mer polymer, $w_N(p)$, can be calculated as (50):

$$w_{N(p)} = Np^{N-1}(1-p)^2 \tag{2}$$

where $p$ is the extent of reaction defined as the ratio of the number of formed physical "bonds" between the 5-bead "monomers" to the maximum possible number of "bonds". **Figure 3B** shows that the simulated weight distribution of the supramolecular assemblies agrees quantitatively well with the Flory theory without any fitting process: the sole parameter $p$ is directly obtained by counting the fraction of reacted sites in the simulations. This agreement confirms that the underlying assumption of a uniform extent of reaction in the Flory theory is valid for self-assembly of the 5-bead model.

To understand the polymerization process, we introduce a reversible kinetic model for supramolecular polymerization:

$$\frac{dp}{dt} = k(1-p)^2 - k_b p \tag{3}$$

where $k$ is the forward polymerization rate constant, and $k_b$ is the bond breaking/disassociation constant. Solving Eq. (2) yields the closed-form time evolution of the extent of reaction:

$$p(t) = \frac{2k}{\sqrt{k_b(4k+k_b)} + 2k + k_b + 2\sqrt{k_b(4k+k_b)}\left(e^{t\sqrt{k_b(4k+k_b)}} - 1\right)^{-1}} \tag{4}$$

which converges at infinite time to the steady-state extent of reaction:

$$p_\infty = \lim_{t\to\infty} p = \frac{2k}{\sqrt{k_b(4k+k_b)} + 2k + k_b}. \tag{5}$$

**Figure S4A** shows that the reversible kinetic model fits quantitatively well with simulation results of time dependent extent of polymerization reaction, $p(t)$. Moreover, the steady-state extent of reaction is $p_\infty = 0.43, 0.73, 0.94$, respectively, for the pole charges of $q = \pm 3.8\ e, \pm 4.0\ e, \pm 5.0\ e$, confirming the pole charges as an effective tuning parameter for supramolecule polymerization.

However, the tunability of polymerization through pole charge is restricted to a narrow range of pole charge magnitudes. As summarized in **Figure 3F**, when $q \leq 3e$, no stable assemblies form; within the intermediate regime of $3e \leq q \leq 5e$, the extent of polymerization reaction is continuously tunable. In this regime, the time-dependent extent of reaction deviates significantly from predictions of the traditional irreversible kinetic model of polymerization suitable for strong covalent bonds, while being well captured by our reversible kinetic model, supporting the existence of reversible supra-molecular polymerization that's tunable by physical interactions.

When $q \geq 5e$, polymerization becomes consistently extensive; Notably, even at a large pole charge of $\pm 13.0\ e$, the predicted equilibrium extent of reaction remains below unity ($p_\infty = 0.943$) (**Figure S5A**), resulting in a broad chain length distribution with most probable weight fraction close to 20 monomers (**Figure S5B**). This resulting filament length closely matches the experimentally observed order of magnitude and reproduces the broad filament length distributions observed for PB1 assemblies (**Figure 2C**). The saturation of $p_\infty$ at large $\pm q$ below 1 may originate from electrostatic screening encoded in the minimal coarse-grained simulation model.

Moreover, the average persistence length for filaments formed by monomers with $q = \pm 13\ e$ and $d_{np} = 2.24$ nm is about 100 nm (**Figure S6**), same order of magnitude as experimental measurements.

Overall, the agreement of time-dependent extent of reaction, $p(t)$, with the reversible kinetic model corroborates the hypothesis that the supramolecular self-assembly is tunable by physical interactions driven by polar protein domains. The identification of an intermediate range of $3e \leq q \leq 5e$ establishes the pole charge as an effective control parameter to tune supramolecular self-assembly by protein structure. Although this region of pole charge is narrow, the onset of polymerization remains diffuse rather than abrupt. This is consistent with the model assumption of screened electrostatic interactions and the predominantly one-dimensional assembly topology, where finite-temperature sharp transitions are generally suppressed for short-range interactions in one-dimensional systems (51, 52). Finally, the saturation of $p_\infty$ when $q \geq 5e$ indicates the robustness of protein polymerization at larger pole charge magnitudes.

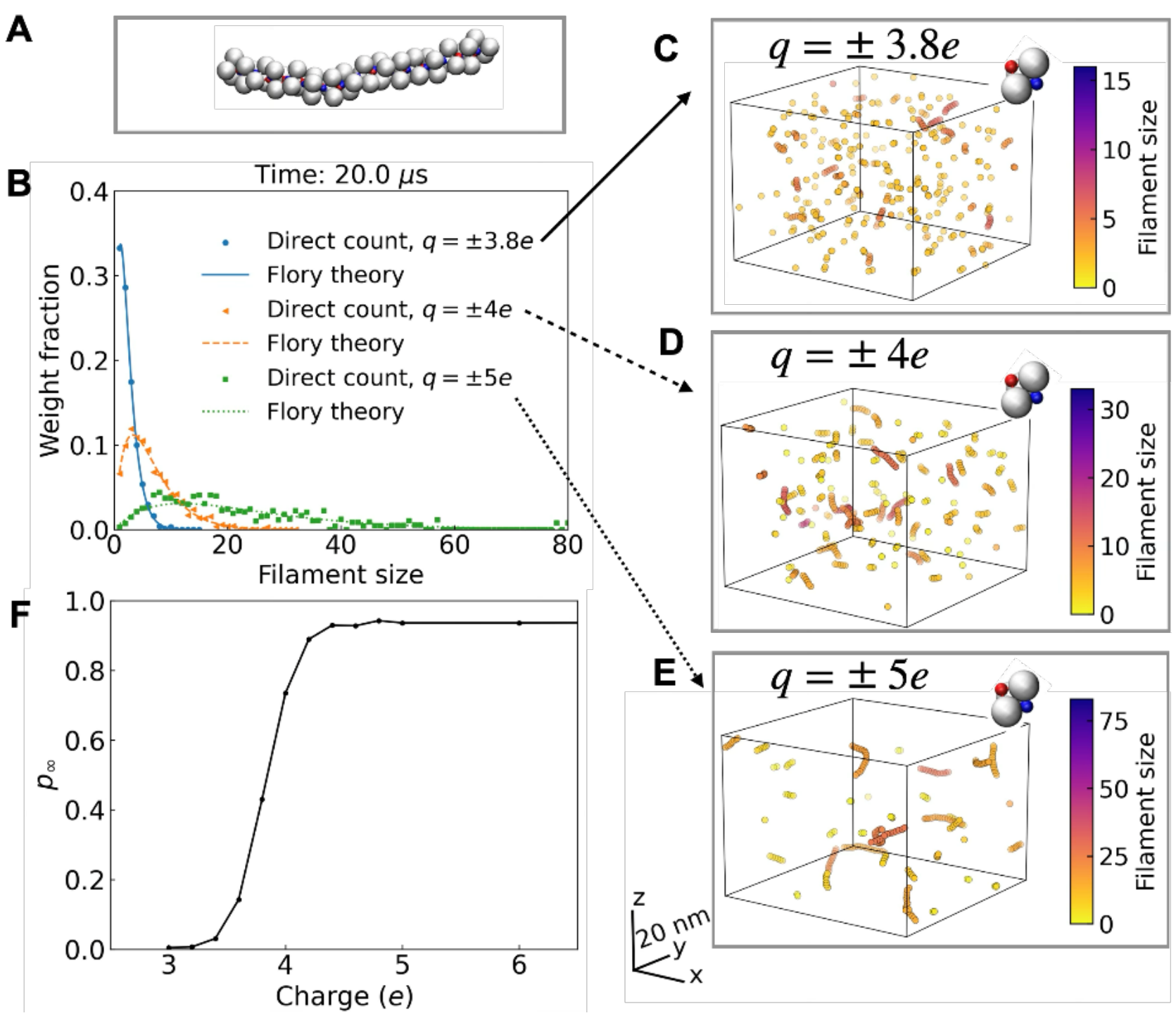

**Figure 3. Self-assembly of the 5-bead model is tunable in a narrow pole-charge regime. A)** Zoomed-in snapshot of a filament formed by 5-bead model. **B)** Weight fraction of supramolecular structures polymerized at 20 $\mu s$ by the 5-bead models with $q = \pm 3.8\ e$ (blue circles), $q = \pm 4.0\ e$ (orange triangles), and $q = \pm 5.0\ e$ (green squares) as a function of the filament size, together with the corresponding predictions of Flory theory (solid blue, dashed orange and dotted green curve, respectively). The zoomed-in snapshots of supramolecular structures self-assembled at 20 $\mu s$ by the 5-bead models with pole charges of **C)** $q = \pm 3.8\ e$, **D)** $q = \pm 4.0\ e$, and **E)** $q = \pm 5.0\ e$. The colorbars indicate the filament size in terms of number of monomers per filament. **F)** the steady-state extent of reaction as a function of pole charge for a fixed $d_{np} = 2.24$ nm.

### Overall protein architecture and rotation flexibility regulate the polymerization structures and dynamics

To understand the role of the remaining protein architecture other than the major polar structured domain, we explore the self-assembly of the 6-bead model. With the parameters for the 5-bead domain fixed at $d_{np} = 2.24$ nm and $q = \pm 13e$, the 6-bead model introduces two additional parameters: $d_{others}$ and $k_{angle}$, as defined previously in the model description.

First, to explore the structural openness of the full-length protein, we vary $d_{others}$ to be 3 nm, 5 nm, and 9 nm, corresponding to progressively more compact to more open six-bead models. Unless otherwise specified, we choose $k_{angle}$ to be 0 to represent a flexible linkage between the polar structured domain (5-bead domain) and the remainder of the protein, allowing the five-bead domain to freely rotate relative to the larger bead.

**Figure 4A** shows that the 6-bead model assemble to elongated structures that are thicker and shorter than filaments assembled by 5-bead model. **Figures 4C to D** show snapshots of assemblies formed by the 6-bead model after 20 $\mu s$ of simulations for $d_{others} =$ 3nm, 5 nm, and 9 nm, respectively. Compared with the long filaments assembled by the 5-bead model under the same pole charge and pole-to-pole distance (**Figures 3C to E**), the 6-bead model forms 1 to 2 orders of magnitude shorter structures at the same simulation time (**Figures 4C to E**), indicating a strong suppressive effect of the additional excluded volume. This suppressive effect is most pronounced for the most compact 6-bead model with $d_{others} = 3$ nm, where dimers dominate the weight distribution of N-mers (**Figure 4B**). Increasing $d_{others}$ progressively relaxes steric constraints and permits longer chains to form: trimers dominate for $d_{others} = 5$ nm and penta-mers become more probable for $d_{others} = 9$ nm (**Figure 4B**).

While the weight distribution of supramolecules formed by the 5-bead model closely agrees with the Flory theory throughout the simulations (**Figure 3B**), assemblies formed by the 6-bead model deviates from the Flory theory (**Figure S7A**). To understand this discrepancy between simulations of 6-bead model and Flory theory, we revisit the self-assembly mechanisms of the 6-bead model mimicking a full-length multidomain protein. For the 5-bead model, a single association probability is sufficient to describe the polymerization process, consistent with uniform reactivity along the growing chain. However, uniform reactivity does not necessarily apply to full-length Bem1 assemblies. The 6$^{th}$ bead partially shields the attractive interactions between polar structured domains embedded within the full-length protein. Because this shielding effect is conformation-dependent, the effective association probability becomes nonuniform along the chain. Therefore, multiple probabilities are required to accurately account for the heterogeneous and spatially modulated shielding effects.

Assuming the shielding effect only affects the nearest neighbors, we define two key parameters to model the polymerization by full-length Bem1: $p_1$ and $p_2$, where $p_1$ is the probability to initialize polymerization of a multimer, and $p_2$ is the probability to extend an existing multimer. The weight fraction can be analytically expressed as (more details in **Supplemental Information**):

$$\begin{cases} w_1 = \dfrac{(1-p_1)(1-p_2)}{1+p_1-p_2} \\ w_N = \dfrac{N p_1 p_2^{N-2}(1-p_2)^2}{1+p_1-p_2} \end{cases}$$

Indeed, **Figure 4B** and **Figure S7A** show that the two-probability model agrees quantitatively well with the simulation data of the most compact 6-bead model ($d_{others} = 3$ nm), without any fitting parameters, as both $p_1$ and $p_2$ are directly measured from the simulations. More open 6-bead models ($d_{others} = 5$ nm and $d_{others} = 9$ nm) correspond to theoretical models with more probability hierarchy (3-p and 4-p, respectively).

To further understand the non-Flory polymerization kinetics and extrapolate to their steady state values, we construct a reversible kinetic model in terms of $c_1$, $c_2$ and $p_2$ (53) (see details in the **Supplemental Information). Figure S7B** shows that the kinetic theory predicts that both $p_1$ and $p_2$ converge beyond the initial stage for the flexible compact 6-bead model, with the steady state values $p_1(\infty) \approx 0.93$, and $p_2(\infty) \approx 0.39$, shifting the steady-state weight distribution towards substantially shorter chains relative to the 5-bead model (**Figure 4F**), resulting in self-limiting self-assembly for the most compact 6-bead model.

Second, to examine the role of rotational flexibility between the 5-bead domain and the 6$^{th}$ larger bead, we imposed a rigid angle bending stiffness with $k_{angle} = 300\ kcal\ mol^{-1}$. **Figure 4F** shows that while the rigid-rotating 6-bead model also assembles into shorter chains than the 5-bead model, it forms much longer chains than the flexible-rotating 6-bead model. This result looks surprising at first glance, as the rigid 6th bead is expected to pose a harder barrier to associate adjacent monomers. However, the freely rotating 6th bead shields a larger portion of the 5-bead domain, and thus entropically favors the shielded unbonded state. This resembles the recently identified entropic barrier created by intrinsically disordered regions, which can hinder the entrance of "intruders" (23). Consequently, rotational flexibility is identified as a key factor to inhibit the supramolecular self-assembly.

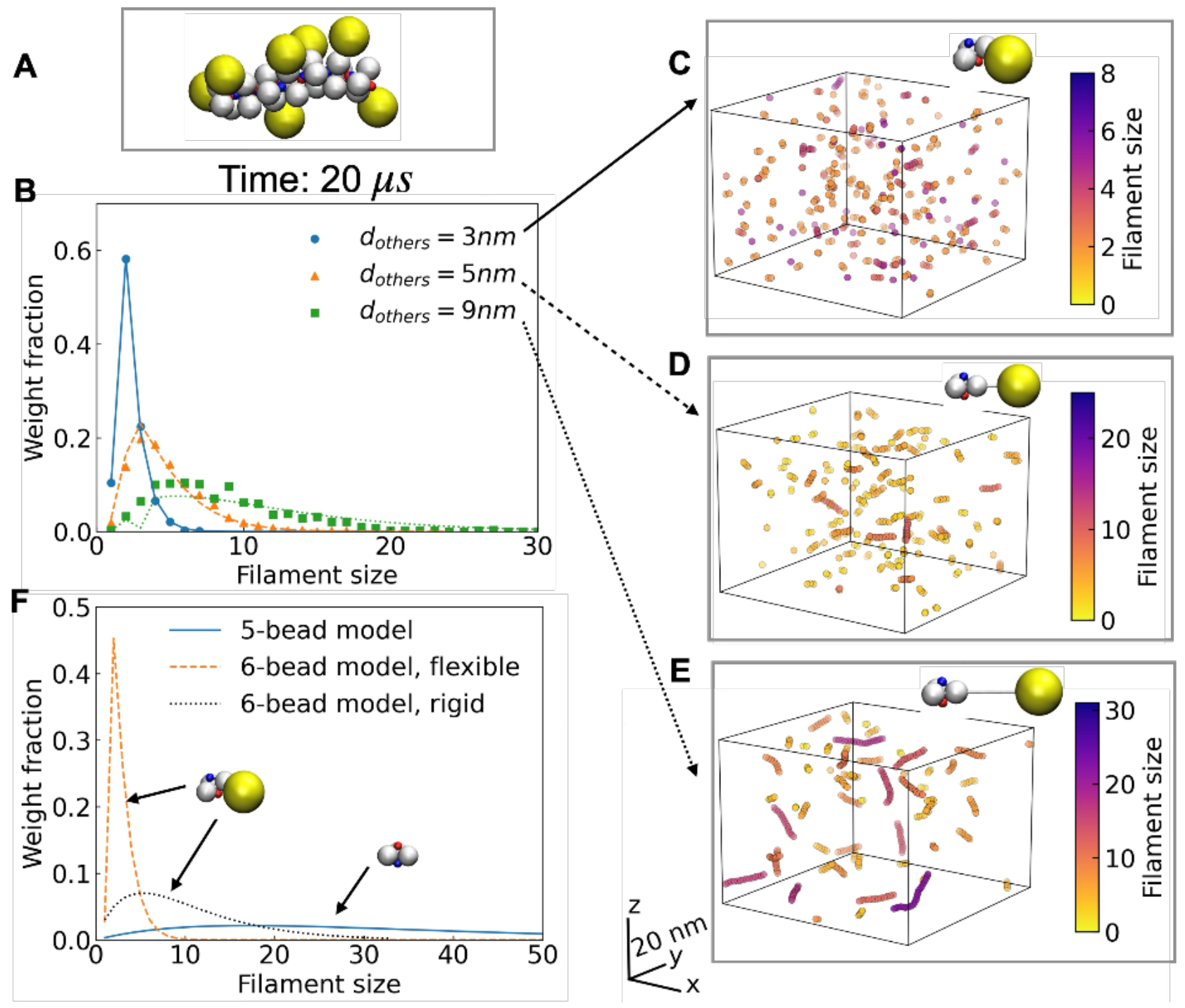

**Figure 4**. **Internal distance and rotational stiffness regulate self-assembly size in the 6-bead model. A)** Zoomed-in snapshot of a filament formed by 6-bead model. **B)** Weight fraction distribution of N-mers at *20* $\mu s$ assembled by the 6-bead models with $d_{np} = 2.24$ nm, $q = \pm 13e$ and $d_{others} = 3$ nm (blue circles), $d_{others} = 4$ nm (orange triangles) and $d_{others} = 9$ nm (green squares). Zoomed-in snapshots of supramolecular structures self-assembled by the 6-bead models with $d_{np} = 2.24$ nm, $q = \pm 13e$ and **C)** $d_{others} = 3$ nm, **D)** $d_{others} = 4$ nm and **E)** $d_{others} = 9$ nm. **F)** Steady-state weight fraction distribution of N-mers for the 5-bead model with $d_{np} = 2.24$ nm and $q = \pm 13e$ (blue solid curve), that for the 6-bead model with $d_{np} = 2.24$ nm, $q = \pm 13e$ and $d_{others} = 3$ nm that allows flexible rotation between the 5-beads and the larger bead (orange dashed curve) and that for the same 6-bead model that does not allow flexible rotation between the 5-beads and the larger bead (black dotted curve).

## Conclusion

In conclusion, we show that the multidomain cell polarity protein Bem1 can self-assemble into supramolecular structures through a mechanism controlled by charge polarity and protein architecture. Our *in vitro* experiments identify PB1 domain as the primary driver for Bem1 self-assembly: isolated PB1 forms long, thin filaments, whereas full-length Bem1 forms thicker and substantially shorter structures, indicating that the remaining regions regulate continued self-assembly. Electrostatic analysis reveals that PB1 possesses a highly heterogeneous and strongly polar surface charge distribution.

Minimal coarse-grained simulations reproduce these behaviors and provide a physical interpretation. A charge-polar domain model undergoes reversible chain polymerization with chain-length distributions quantitatively consistent with Flory theory, indicating approximately uniform reactivity along the filament. In contrast, embedding the same domain within a larger flexible architecture generates systematic deviations from Flory behavior due to steric shielding and conformation-dependent accessibility.

These effects can produce self-limited self-assembly, despite unchanged attractive interactions introduced by the charge-polar structured domain. We note that the present kinetic framework is limited to step-growth polymerization through monomer addition. On longer timescales, agglomeration or chain–chain association processes may become relevant, as occasionally observed in the simulations. These secondary assembly pathways may contribute to the rare appearance of exceptionally long assemblies in experiments, even for full-length Bem1. We also neglect interactions associated with the unstructured regions for simplicity, although these contributions may become important under different protein architectures or environmental conditions.

Prior reports of smaller Bem1 multimers and membrane-bound nanoclusters *in vivo* (40, 54) further suggest that interacting partners and cellular context may regulate assembly size and dynamics for signaling process. Further investigation of the (self-)assembly kinetics of signaling proteins like Bem1 may help elucidate the transient and responsive dynamics that emerge within the complex, high density environment of signal establishment.

More broadly, our results suggest that biomolecular self-assembly of multidomain proteins can be governed by two complementary design principles: sequence-encoded interactions and architecture-encoded geometric regulation. Together, these perspectives may enable predictive understanding of how multidomain proteins assemble into condensates, filaments, and other supramolecular structures (55). Integration of our experimental results and architectural minimal model framework with sequence-based sticker-spacer framework may provide a promising route towards predictive and *de novo* design of biomolecular assemblies and provides insight into the role of supramolecular assemblies in cellular signaling processes.

## Materials and Methods

### Sequence and structure analysis.

Protein and domain information was obtained from the the Uniprot database(48) and the InterPro (56) database respectively. The predicted protein structure for Bem1 was obtained from Alphafold (AF-P29366-F1-v6) (57, 58). Electrostatic calculations were performed with the Adaptive Poisson-Bolzmann Solver (APBS) in Pymol (44). For a detailed description see SI Appendix, section 1.

### Protein purification

The sequence for full length Bem1 was obtained from the genome of Saccharomyces cerevisiae strain W303 transformed into competent competent Dh5$\alpha$ and BL21 DE::3 pLysS cells. Proteins were expressed in Bl21::DE3 pLysS cells and purified through His-affinity chromatography and size exclusion chromatography. Proteins were analyzed on SDS-PAGE and through anti-His Western blotting. For a detailed description of the protein purification, see SI Appendix, sections 2-5.

### Protein characterization

SEC-MALS was performed on a high-performance liquid chromatography (HPLC) unit (1260 Infinity II, Agilent) running in series with an online ultraviolet (UV) detector (1260 Infinity II VWD, Agilent), an eight-angle static light scattering detector (DAWN HELEOS 8+, Wyatt Technology) and a refractometer (Optilab T-rEX, Wyatt Technology). Weight-averaged molar masses were calculated using ASTRA software (Wyatt Technology, version 7.3.2). For a detailed description see SI Appendix, section 6.
Proteins were stained with 2% uranyl acetate solution and imaged on a JEM 1400Plus TEM (JEOL) operated at 120 kV and recorded on a bottom-mounted TVIPS F416 complementary metal-oxide-superconductor camera. For a detailed description see SI Appendix, section 7
Filament length and filament persistence length was determined from nsEM micrographs using EasyWorm (59). For a detailed description see SI Appendix, section 8.

**Simulation set up**

In all MD simulations, 10000 monomers are randomly scattered in a 500 nm by 500 nm by 500 nm box with periodic boundary conditions and implicit solvent. Debye- Hückel model was used to describe the electrostatic interactions with a Debye length of 1 nm, corresponding to 100 mM ionic strength. The excluded volume of beads was simulated by a shifted, smoothly truncated 12-6 Lenard-Jones potential, which is effectively repulsive. The sizes of the beads are summarized in Table S5. Langevin dynamics was used to simulate monomer diffusion at a temperature of 300 K. For a detailed description see SI Appendix, section 9-11.

**Author Contributions**

N.M.H : Designed research, performed research, analyzed data, wrote the paper.
M.S: Designed research, performed research, analyzed data, wrote the paper.
F.v.O: Performed research
E.L: Analyzed data
L.L: Designed research, wrote the paper

**Funding**

L. Laan gratefully acknowledges funding from the European Research Council under the European Union's Horizon 2020 research and innovation programme (grant agreement 758132) and funding from the Netherlands Organization for Scientific Research (Nederlandse Organisatie voor Wetenschappelijk Onderzoek) through a Vidi grant (016.Vidi.171.060).
M. Shen and E. Lim are grateful to the funding by Undergraduate Research Opportunity Center (UROC) at California State University, Fullerton (CSUF). E. Lim acknowledges support by the Dan Black Scholarship. This research was supported in part by grant NSF PHY-2309135 and the Gordon and Betty Moore Foundation Grant No. 2919.02 to the Kavli Institute for Theoretical Physics (KITP).

**Acknowledgments**

We thank Ramon van der Valk, Cecilia de Agrela Pinto and Wiel Evers for their experimental assistance and Renske Voerman for her assistance in data analysis. We thank Nynke Dekker (University of Oxford) for the plasmid pET28a-His-mcm10-Sortase-Flag. We thank Kurt Kremer, Alexander Grosberg, Fyl Pincus and Arjen Jakobi for their helpful discussions. We thank Chloe Waters for reading the manuscript and providing helpful feedback.

Supporting Information for

# Tunable supramolecular polymerization from charge heterogeneity and protein architecture

Nynke Marije Hettema[†1], Meng Shen[†2*], Frank van Opstal[1], Emmett Lim[2], Liedewij Laan[1*]

[†] These authors contributed equally to this work

[1] Bionanoscience Department, Delft University of Technology, Delft 2629HZ, The Netherlands

[2] Department of Physics, California State University Fullerton, Fullerton, CA 92831, USA

*Corresponding Authors: Liedewij Laan and Meng Shen

**Email:** L.laan@tudelft.nl and meshen@Fullerton.edu

**This SI-document includes:**

Supplemental methods

Figures S1 to S7

Tables S1 to S5

Legends for Movies S1 to S3

SI References

**Other supporting materials for this manuscript include the following:**

Movies S1 to S3

## Supplemental methods

### 1. Sequence and structure analysis

Protein sequences were obtained from the Uniprot database(1) and corresponding domain information was obtained from the InterPro(2) database. The amount of domains per protein was counted from the obtained domains of the InterPro(2) database, where domains overlapping for less than 75% in sequence were considered individual domains. The sum of the sequence covered by domains was divided by full-length protein sequence per protein to determine domain coverage. The predicted protein structure for Bem1 was obtained from Alphafold (AF-P29366-F1-v6) (3, 4). Electrostatic calculations were performed with the Adaptive Poisson-Bolzmann Solver (APBS) in Pymol (5) using the "formal_charge and vdw" method for molecule preparation.

### 2. Plasmid construction

The sequence for full length Bem1 was obtained from the genome of Saccharomyces cerevisiae strain W303 and was amplified through PCR. The target vector (pET28a-His-mcm10-Sortase-Flag, based on pBP6 in Douglas and Diffley (6)) was amplified through PCR. Gibson assembly was used to combine the target and protein-sequence. The resulting plasmids were used to transform chemically competent Dh5$\alpha$ and BL21 DE::3 pLysS cells and plated onto a Petri dish with Lysogeny broth agar and a Kanamycin antibiotic marker. The product (pRV009, containing the sequence for His-Bem1-Flag) was truncated to produce the plasmids for His-Bem1$_{1\text{-}464}$-Flag (Bem1ΔPB1) and His-Bem1$_{478\text{-}551}$-Flag (PB1) through Gibson Assembly.

A list of all plasmids and protein constructs used in this publication can be found in Table S1. The construction route of plasmids and the primers used in each PCR are stated in Figure S3 and Table S2. The amino acid sequence of the proteins used in this publication are stated in Table S3.

### 3. Protein expression and purification

All constructs were expressed in Bl21::DE3 pLysS cells. The cells were grown until an OD600 of 0.6 at 37°C in Lysogeny broth supplemented with Kanomycin. After addition of 1.0 mM IPTG to induce expression, the cells for His-Bem1-Flag and His-Bem1$_{1\text{-}464}$-Flag were grown for 18 h at 18°C and the cells for His-Bem1$_{478\text{-}551}$-Flag were grown for 2 h at 37°C. All buffers used for purification are stated in Table S4. Cells were harvested through centrifugation and resuspended in lysis buffer after which they were lysed with a high pressure homogenizer(French press cell disruptor, CF1 series Constant Systems) in 5-10 rounds at 4°C. The cell lysate was centrifuged at 37000x g for 30 min and the supernatant was loaded onto a HisTrap$^{TM}$ excel column (Cytiva). After washing with His-AC washing buffer, the protein was eluted in a linear gradient of His-AC washing and elution buffer. His-Bem1-Flag and His-Bem1$_{1\text{-}464}$-Flag were further purified by size exclusion chromatography on a HiPrep 16/60 Sephacryl S-300 HR (Cytiva) column using SEC buffer. His-Bem1$_{478\text{-}551}$-Flag was dialysed twice in SEC buffer. 10% glycerol was added to the samples which were then flash frozen in liquid nitrogen and stored at -80°C.

### 4. SDS-PAGE

Proteins in the main manuscript were analyzed on freshly prepared SDS-PAGE gels (15% acrylamide). For Figure S1 samples were analyzed on a SurePAGE, Bis-Tris 10% gel (GenScript) for better separation between domain and full-length protein sizes. Protein samples were mixed with SDS loading buffer and kept for 5 min at 95°C. Samples were loaded and run on gel for 5 min at 130 V followed by 55 min at 180 V (PowerPac Basic Power Supply, Bio-Rad). Samples were stained using SimplyBlue SafeStain (Invitrogen) overnight and imaged on a ChemiDoc MP (Bio-Rad) using the Coomassie feature and automatic exposure time determination.

## 5. Western blotting

The sample was transferred from SDS-PAGE to a blotting membrane using the (Trans-Blot Turbo Mini 0.2 µm PVDF Transfer Pack, Bio-Rad) using the 'Mixed MW' program of the Trans-Blot Turbo Transfer System (Bio-Rad). The blotting membrane was incubated with Immobilon signal enhancer (Millipore) at 4°C for 18 h. The blotting membrane was incubated with primary antibody(His Tag Antibody, Mouse (OAEA00010, Aviva Systems Biology)), diluted in Immobilon signal enhancer (dilution: 1:4000) at room temperature for 1 h. It was washed thrice after a 20 min incubation at room temperature with TBS-T (10 mM Tris-HCl (pH=7.5), 150 mM NaCl, 0.1 v/v% Tween-20 (Sigma Aldrich)). The blotting membrane was incubated with secondary antibody (IgG2b Antibody HRP conjugated, Goat Anti-Mouse (OASA06620, Aviva Systems Biology)), diluted in Immobilon signal enhancer, at room temperature for 1 h, after which it was again washed thrice with TBS-T. SuperSignal West Pico Mouse IgG Detection Kit (Thermo Scientific) was used for activation. Imaging was done on a ChemiDoc MP (Bio-Rad) using the 'Chemi Sensitive' feature and automatic exposure time determination. Precision Plus Protein Unstained standard (Bio-Rad) and Precision Plus Protein All Blue Standard (Bio-Rad) were used as protein standards.

## 6. SEC-MALS

Protein samples diluted in SEC buffer to a concentration of 0.5 mg/ml were loaded onto a Superdex 200 increase 10/300 gl connected to a high-performance liquid chromatography (HPLC) unit (1260 Infinity II, Agilent) running in series with an online ultraviolet (UV) detector (1260 Infinity II VWD, Agilent), an eight-angle static light scattering detector (DAWN HELEOS 8+, Wyatt Technology) and a refractometer (Optilab T-rEX, Wyatt Technology).Weight-averaged molar masses were calculated using ASTRA software (Wyatt Technology, version 7.3.2) based on the Rayleigh scattering at different angles and dn/dc=0.185 ml/g. The weight fraction distribution was obtained by plotting the dRI relative concentration against the weight-averaged molar masses, bin sizes were determined by multiples of the expected monomer mass of 69.9 kDa for the full-length Bem1 and 59.5 kDa for Bem1ΔPB1.

## 7. nsEM

Protein was diluted in SEC buffer to 1 µM. 3 µL protein sample was deposited on a glow-discharged carbon coated EM grid (Quantifoil). The grid was incubated with the sample for 1 minute, after which excess sample was removed through side-blotting. The grid was then washed 3 times with water and stained with 3 µl 2%uranyl acetate solution for 30 s, excess stain was removed through side-blotting. All samples were imaged on a JEM 1400Plus TEM (JEOL) operated at 120 kV and recorded on a bottom-mounted TVIPS F416 complementary metal-oxide-superconductor camera.

## 8. Filament characterization

Filament length and filament persistence length was determined from nsEM micrographs using EasyWorm (7). Within this Matlab based software tool, filaments were manually traced, after which the squared end-to-end distance and the filament contour were used to fit a worm-like chain model to determine the persistence length. Given the worm-like chain model assumption underlying this calculation, chains longer than 150nm (roughly twice the estimated persistence length) were omitted from fitting.

The weight distribution of filaments was calculated using

$$w_N = \frac{n_N N}{\sum_{N'} n_{N'} N'},$$

where $N$ is the number of monomers in each polymer and $n_N$ the number of N-mers. N was estimated by dividing the contour length of each filament in nsEM by the expected monomer size of 3nm under the assumption of a spherical folded domain.

## 9. Force field parameters of minimal coarse-grained molecular dynamics (MD) simulations

All simulations are performed in Large-Scale Atomic/Molecular Massively Parallel Simulator (LAMMPS) (8). The size of the three central beads in the 5-bead model is parameterized such that the effective cross-sectional diameter of the resulting supramolecular chains matches the experimentally observed filament diameter of 3nm. The even larger neutral bead in the 6-bead model is twice larger than the central beads. More detailed parameters for the shifted smoothly truncated Lennard-Jones potential are listed in Table S5. Lorentz-Berthelot (LB) mixing rules are adopted for cross-species interactions.

## 10. Hierarchical probability model and corresponding kinetics model

Assuming the shielding effect only affects the nearest neighbors, two key parameters can be defined to model the polymerization of the most compact 6-bead model ($d_{others} = 3$ nm): $p_1$and $p_2$, where $p_1$is the probability to initialize polymerization of a multimer, and $p_2$ is the probability to extend an existing multimer. We define $c_N$ as the concentration of N-mers, then

$$\begin{cases} c_1 = A(1-p_1)^2 \\ c_{N\geq 2} = A(1-p_1)p_1 p_2^{N-2}(1-p_2) \end{cases}$$

Defining $\pi_N = \frac{c_N}{c_{tot}}$ as the normalized concentration, we obtain:

$$\begin{cases} \pi_1 = 1-p_1 \\ \pi_{N\geq 2} = p_1 p_2^{N-2}(1-p_2) \end{cases}$$

Therefore, $p_1$ can be directly determined from $\pi_1$, and $p_2$ can be estimated by an average ratio of $\frac{c_{N+1}}{c_N}$. Similarly, the weight fraction can be analytically expressed as:

$$\begin{cases} w_1 = \dfrac{(1-p_1)(1-p_2)}{1+p_1-p_2} \\ w_N = \dfrac{N p_1 p_2^{N-2}(1-p_2)^2}{1+p_1-p_2} \end{cases}$$

To further understand the time evolution of this non-Flory polymerization extrapolate to their steady state values, we also construct a reversible kinetic model in terms of $c_1, c_2$ and $p_2$ (9):

$$\begin{cases} \dfrac{\mathrm{dc_1}}{dt} = -2\mathrm{k_1c_1^2} + 2\mathrm{k_{b1}c_2} - \dfrac{2\mathrm{k_2c_1c_2}}{1-p_2} + \dfrac{2\mathrm{k_{b2}c_2}p_2}{1-p_2} \\ \dfrac{\mathrm{dc_2}}{dt} = \mathrm{k_1c_1^2} - \mathrm{k_{b1}c_2} - 2\mathrm{k_2c_1c_2} + 2\mathrm{k_{b2}c_2}p_2 \\ \dfrac{dp_2}{dt} = 2k_2c_1 - 2k_{b2}p_2 - \dfrac{k_1 p_2 c_1^2}{c_2} + k_{b1}p_2 \end{cases}$$

This is a system of coupled ordinary differential equations (ODE) with no analytical solutions. Therefore, we fit numerical solutions of the above ODEs to the available simulation data to determine the four coefficients $k_1, k_{b1}, k_2, k_{b2}$. These parameters are then used to solve for the steady-state (and equilibrium) concentrations together with the closure relation $\sum_N \quad c_N = c_{tot}$.

For more open 6-bead models ($d_{others} > 3\ nm$), more different probabilities are necessary to agree between simulations and theory.

## 11. Persistence length calculation from simulations

Assuming free azimuthal rotation for the "bonds", the correlation between the "bond" vectors decay exponentially with the internal distance (10):

$$\langle cos\ cos\ \theta_{\ j-i} \rangle = e^{-\frac{l_b |j-i|}{l_p}}$$

where $l_b$ is the average bond length that can be extracted from the simulations, and $i$ and $j$ are bond indices and $l_p$ is the persistence length.

**Figures**

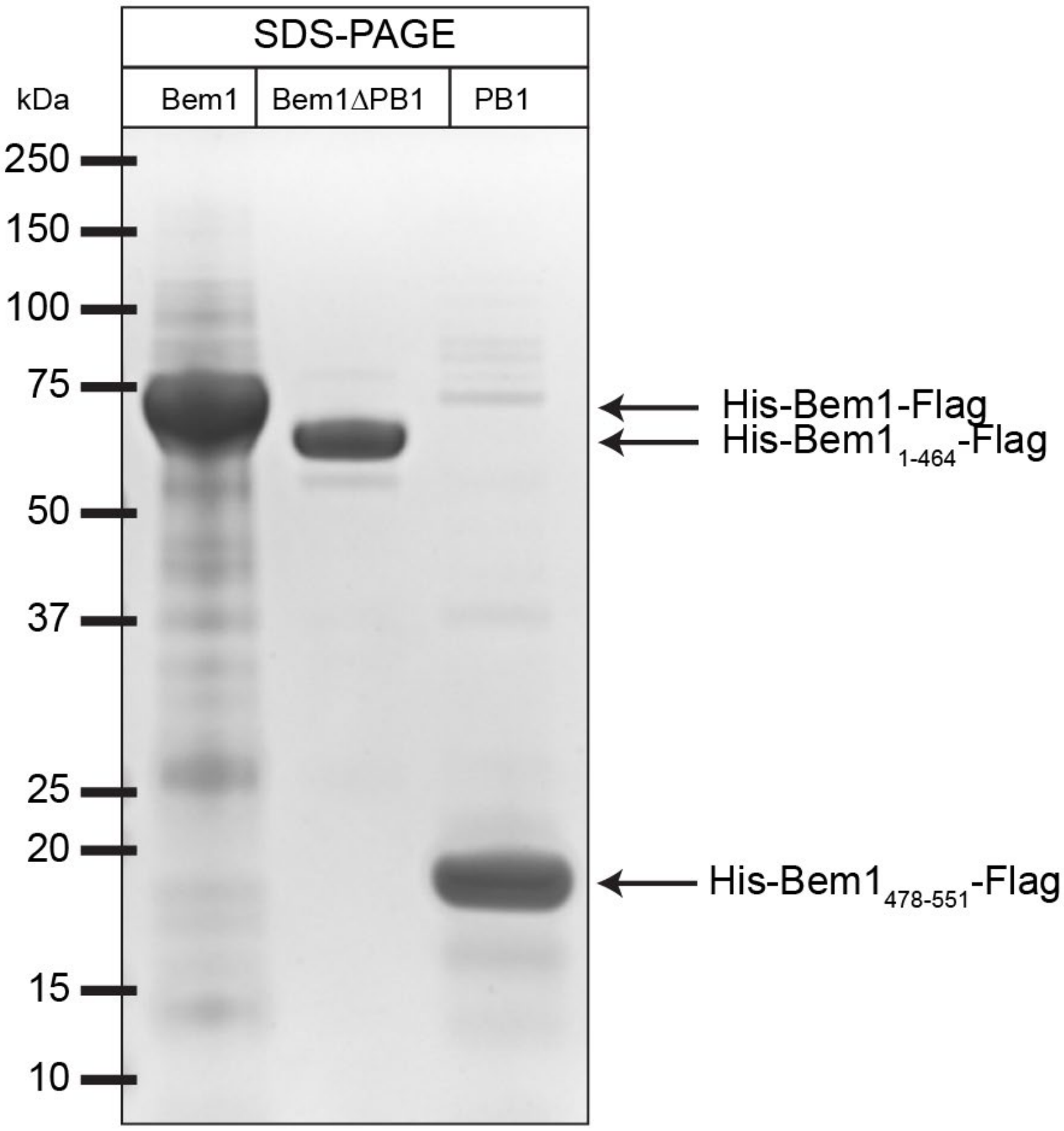

**Figure S1.** SDS-PAGE of protein constructs used in this manuscript. Expected running height indicated by arrows, molecular size Bem1 69.9 kDa, Bem1ΔPB1 59.5 kDa. PB1 17.2 Da.

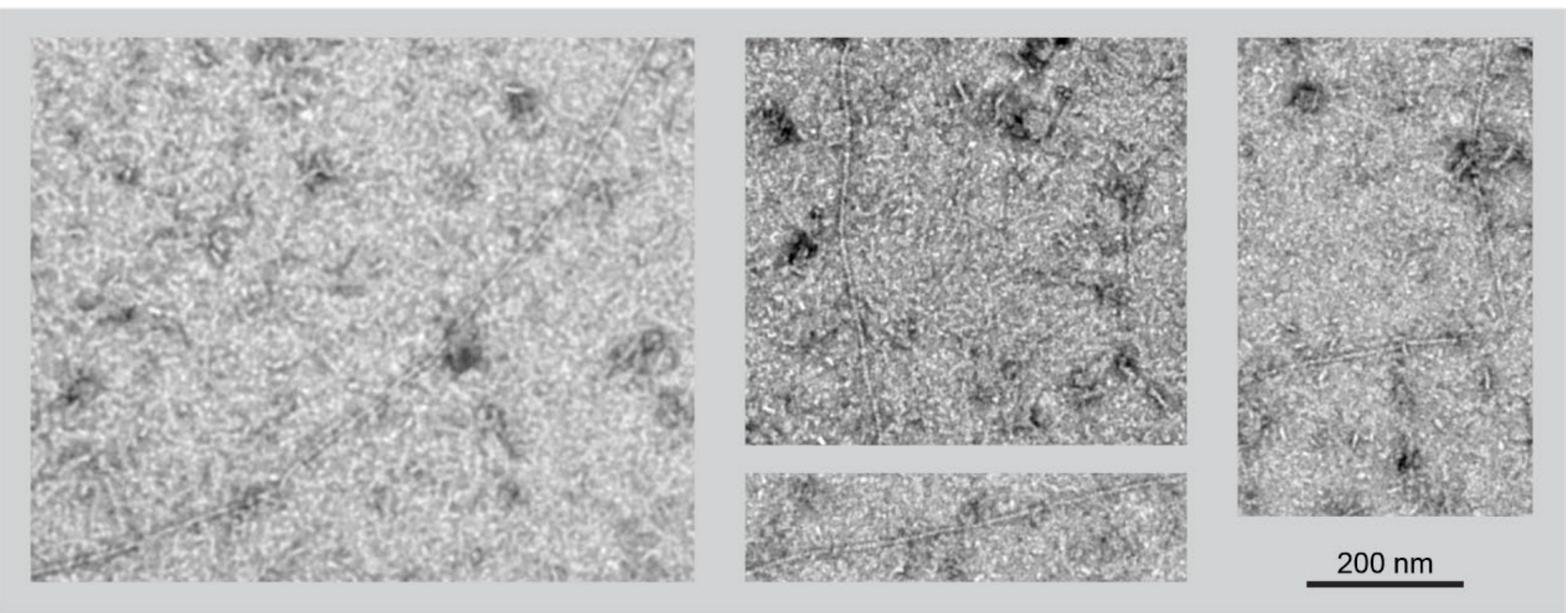

**Figure S2.** Negative-stained transmission electron micrographs of Bem1$_{478\text{-}551}$(PB1 domain) under low salt conditions (25mM Tris (pH=7.5), 50mM NaCl, 5mM MgCl2, 0,5mM 2-MercEtOH). Scalebar is 200nm.

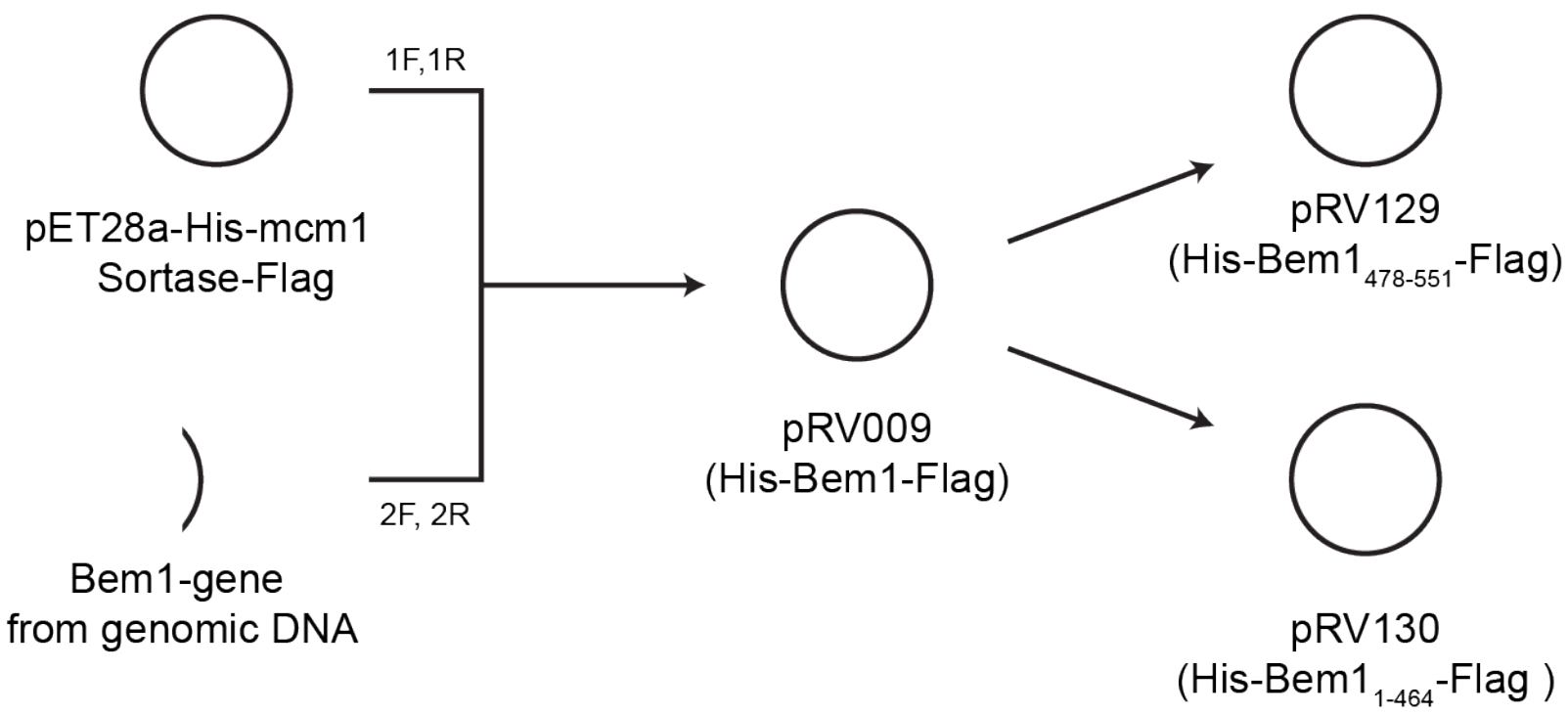

**Figure S3.** Schematic of plasmid construction for all protein constructs used in this publication.

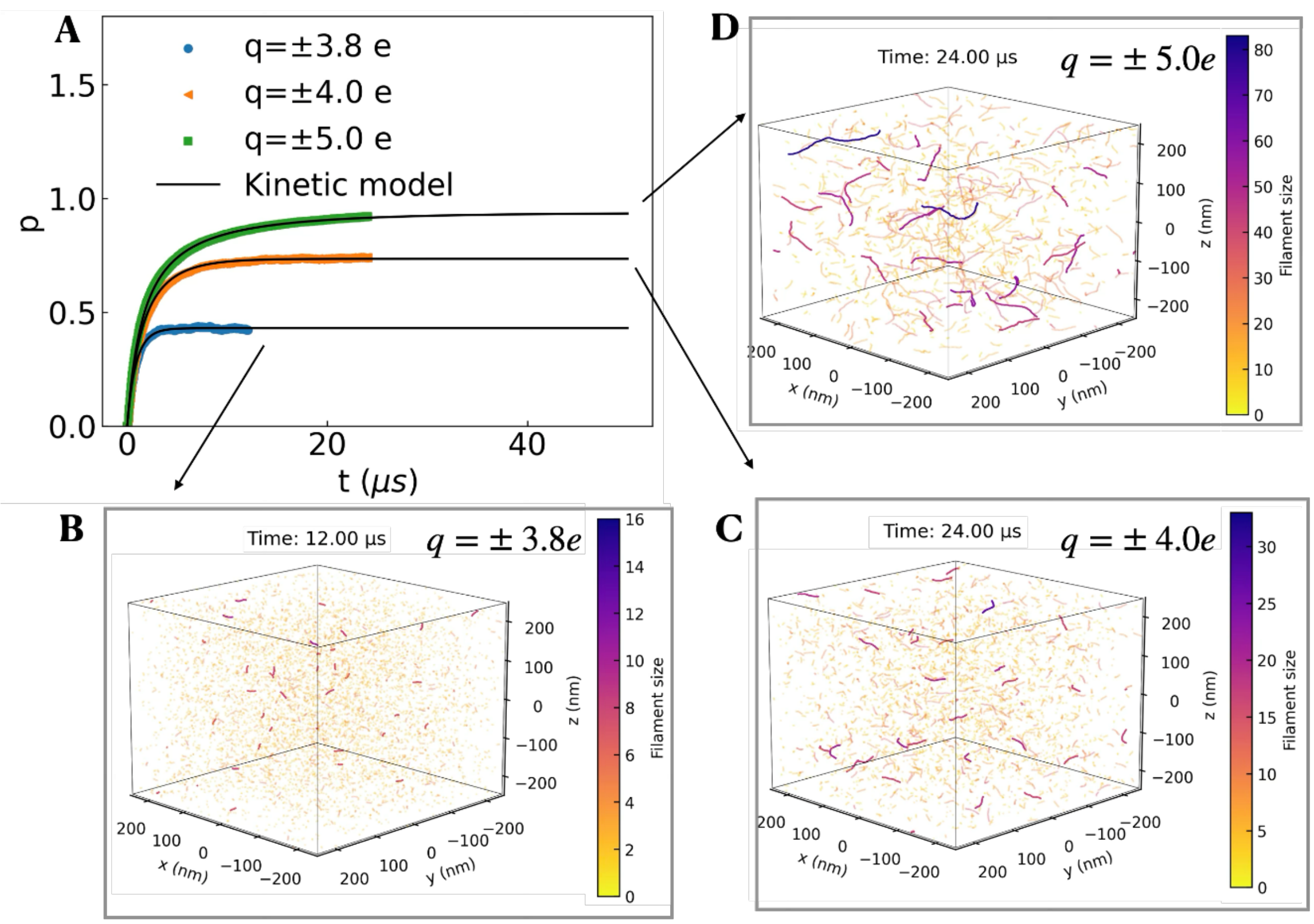

**Figure S4. A)** The time-dependent extent of "polymerization" reaction of the 5-bead models with $q = \pm 3.8\ e$ (blue circles), $q = \pm 4.0\ e$ (orange triangles). The solid curves show fitting and long-time extrapolations using the analytical solution of the reversible kinetic model assuming a uniform reaction rate. The snapshot of whole simulation box **B)** at 12 $\mu s$ for $q = \pm 3.8\ e$, and at 24 $\mu s$ for **C)** $q = \pm 4.0\ e$ and **D)** $q = \pm 5.0\ e$.

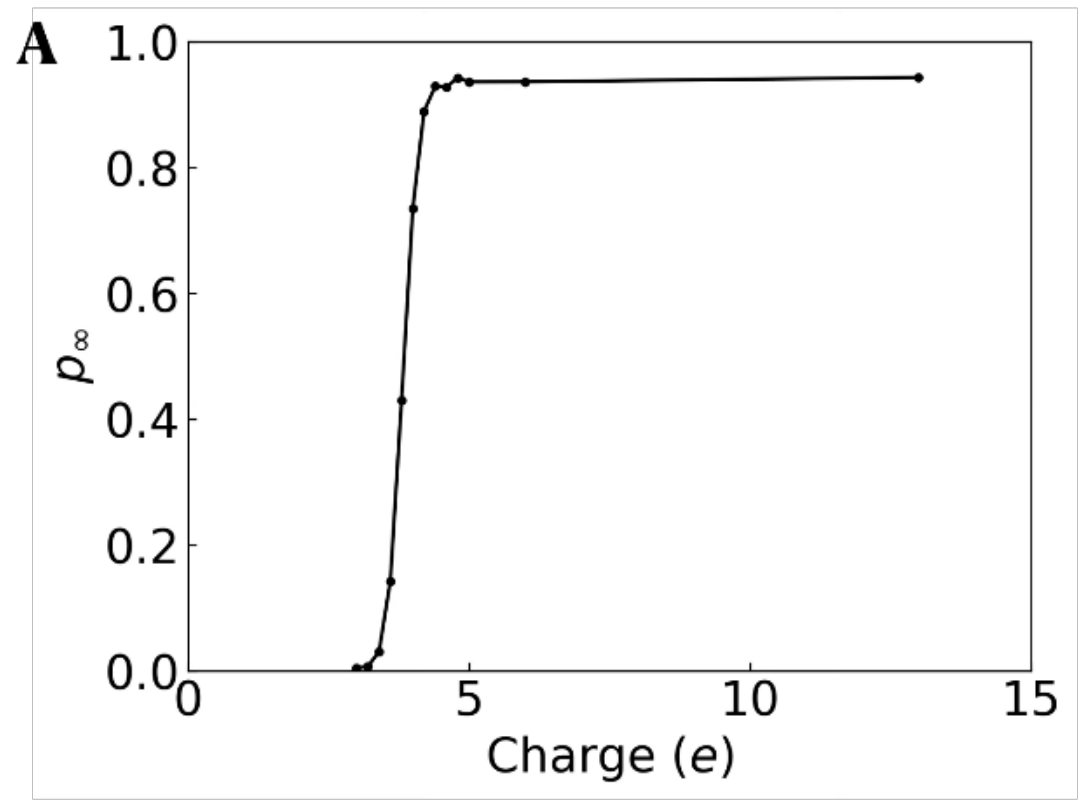

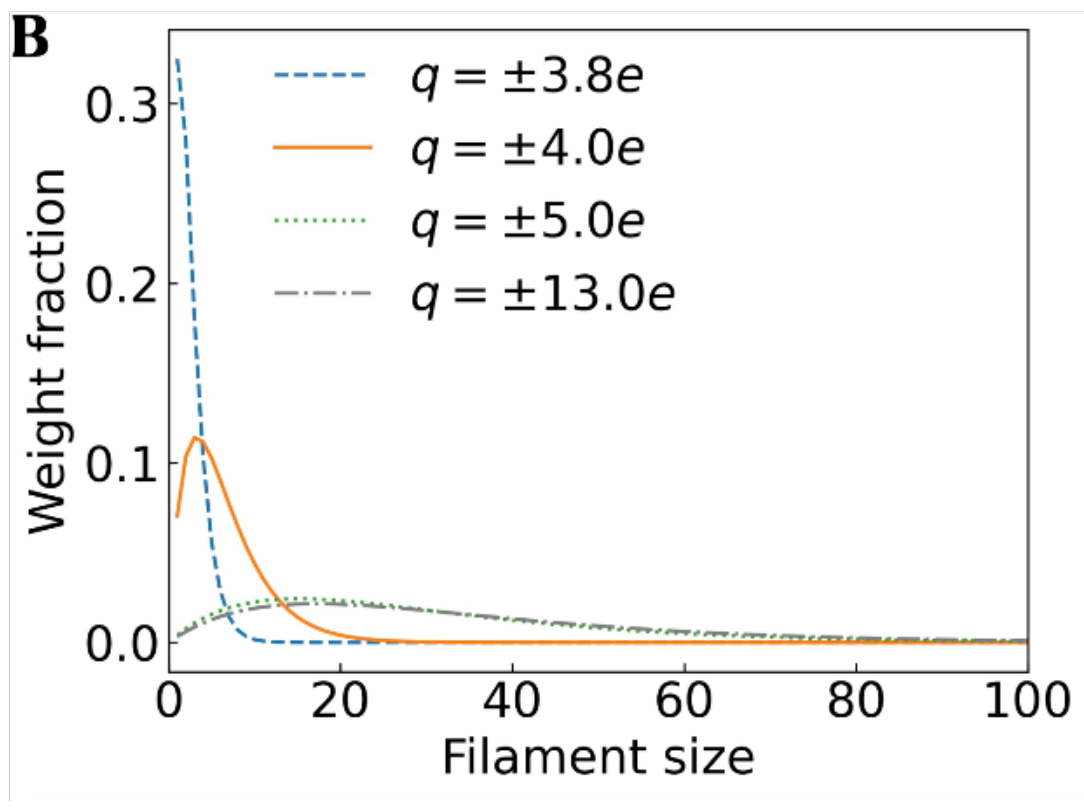

**Figure S5: A)** the steady-state extent of reaction as a function of pole charge for a fixed $d_{np} = 2.24$ nm. **B)** The kinetic-model-predicted steady-state weight fraction of supramolecular structures polymerized by the 5-bead models with $q = \pm 5.0\ e$ (blue circles) and $q = \pm 13.0\ e$ (orange triangles) as a function of the filament size.

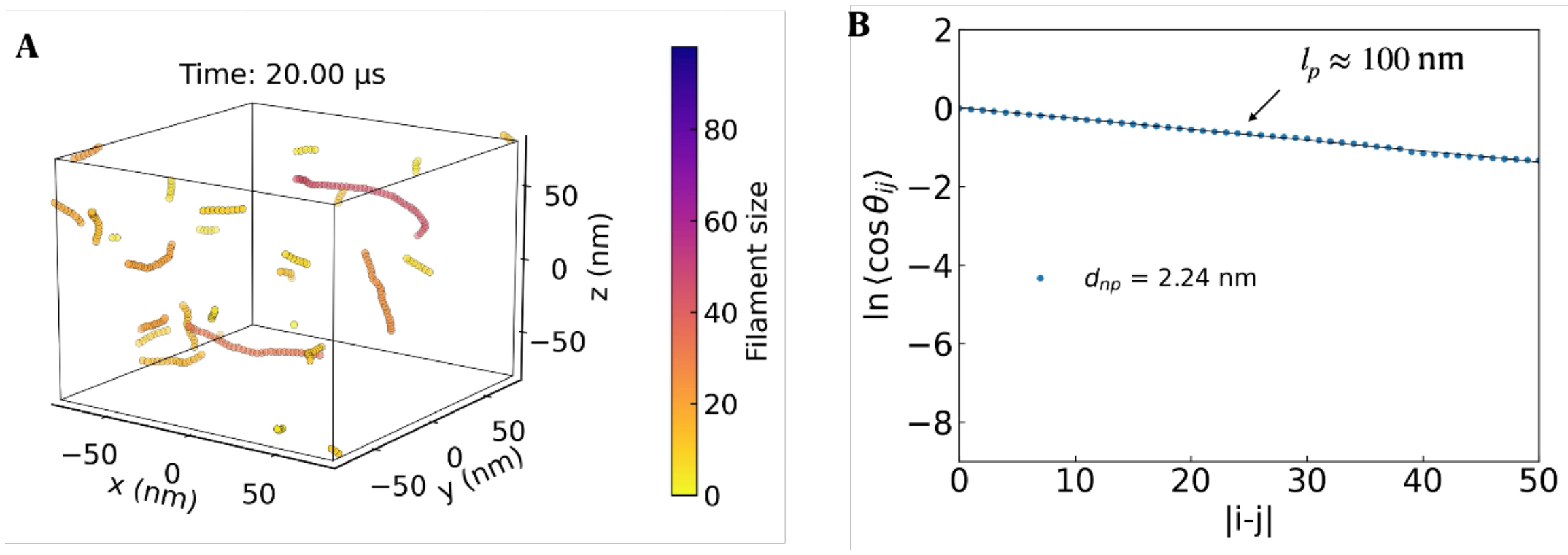

**Figure S6. A)** Zoomed-in self-assembled structures of 5-bead model models with $d_{np} = 2.24\ nm$. **B)** Log of the bond vector correlation as a function of intra-chain indices for various $d_{np}$.

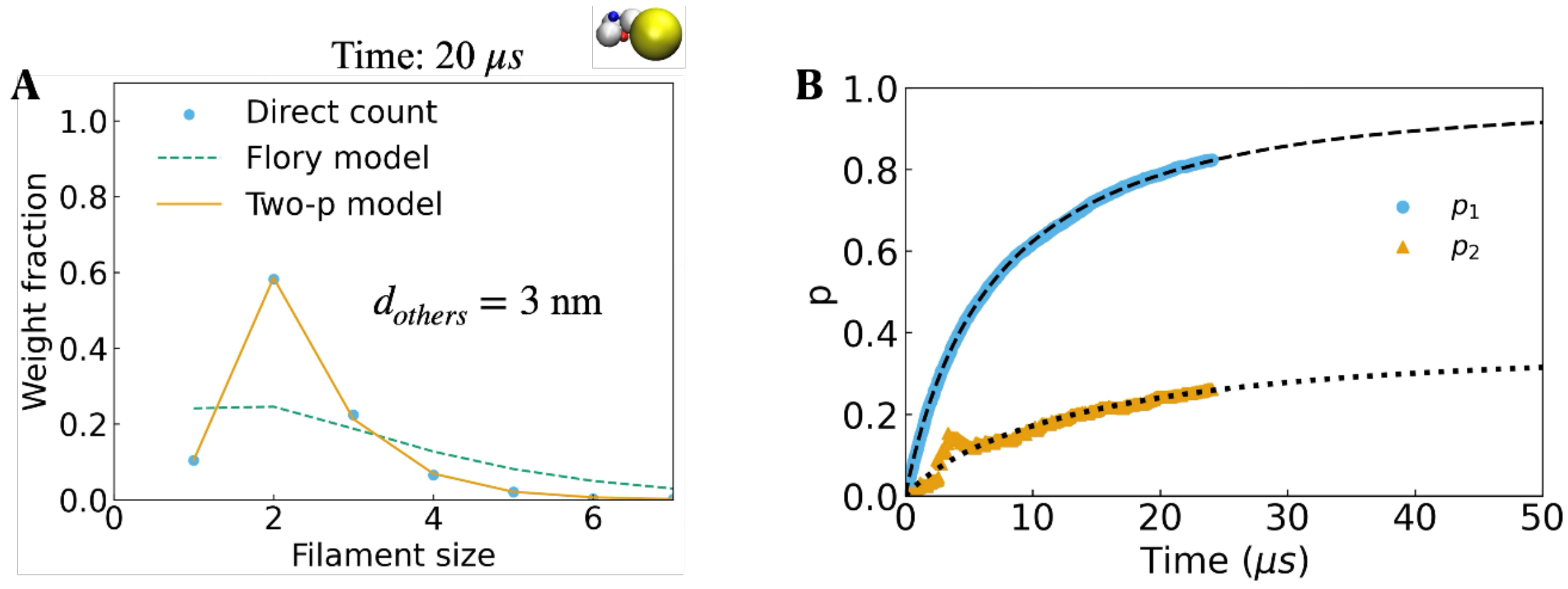

**Figure S7. A)** The weight fraction distribution of N-mers for the 6-bead model with $d_{np} = 2.24$ nm, $q = \pm 13e$ and $d_{others} = 3$ nm from simulations (blue dots), the Flory model prediction (green dashed curve) and two-probability model prediction (orange solid curve). **B)** The time dependent $p_1$ and $p_2$ from simulations (blue circles and orange triangles, respectively) and from the two-probability kinetic model fitting and extrapolation (dashed and dotted curves, respectively).

## Tables

**Table S1.** Plasmids and protein constructs used throughout this publication.

| **Plasmid** | **Description** | **Source** |
|---|---|---|
| pET28a-His-mcm10-Sortase-Flag | Template for pRV009 | Received from N. Dekker (TU Delft), based on pBP6 in Douglas and Diffley |
| pRV009 | His-Bem1-Flag | This work |
| pRV129 | His-Bem1$_{478\text{-}551}$-Flag | This work |
| pRV130 | His-Bem1$_{1\text{-}464}$-Flag | This work |

**Table S2.** Primer overview.

| **Primer** | **Sequence** |
|---|---|
| 1F | GGTGGCGGAGGAAGCT |
| 1R | TTTATCGTCGTCATCgaattcggatcc |
| 2F | caGcaaatgggtcgcggatccgaattcGATGACGACGATAAAATGCTGAAAAACTTCAAACTCTCAAAAAGAGA |
| 2R | CTTGTAGTCACCGCCGGTTTCCGGTAAGCTTCCTCCGCCACCGAGTCTAATATCGTGAACGGAAATTTTCAGT |

**Table S3.** His-Bem1-Flag Protein sequence and sequence features.

| **Feature** | **Sequence** |
|---|---|
| His-Bem1-Flag | MGSS*HHHHHH*SSGLVPRGSH*MASMTGGQQMGRGSEF*DDDDKMLKNFKLSKRDSNGSKG<br>RITSADISTPSHDNGSVIKHIKTVPVRYLSSSSTPVKSQRDSSPKNRHNSKDITSPEKVIKAKYSY<br>QAQTSKELSFMEGEFFYVSGDEKDWYKASNPSTGKEGVVPKTYFEVFDRTKPSSVNGSNSSS<br>RKVTNDSLNMGSLYAIVLYDFKAEKADELTTYVGENLFICAHHNCEWFIAKPIGRLGGPGLVP<br>VGFVSIIDIATGYATGNDVIEDIKSVNLPTVQEWKSNIARYKASNISLGSVEQQQQQSITKPQN<br>KSAKLVDGELLVKASVESFGLEDEKYWFLVCCELSNGKTRQLKRYYQDFYDLQVQLLDAFPAE<br>AGKLRDAGGQWSKRIMPYIPGPVPYVTNSITKKRKEDLNIYVADLVNLPDYISRSEMVHSLFV<br>VLNNGFDREFERDENQNNIKTLQENDTATFATASQTSNFASTNQDNTLTGEDLKLNKKLSDL<br>SLSGSKQAPAQSTSGLKTTKIKFYYKDDIFALMLKGDTTYKELRSKIAPRIDTDNFKLQTKLFDG<br>SGEEIKTDSQVSNIIQAKLKISVHDIRLGGGGSLPETGG_DYKDHDGDYKDHDIDYKDDDDK_* |
| 6His-tag | *HHHHHH* |
| Flag-tag | *DYKDHDGDYKDHDIDYKDDDDK* |
| Thrombin cut site | LVPRGS |
| Enterokinase cut site | DDDDK |
| Sortase recognition motif | LPETG |
| T7 tag | *MASMTGGQQMGRGSEF* |

**Table S4.** Overview of buffers used in protein expression and protein characterization.

| **Buffer** | **Composition** |
|---|---|
| Lysis buffer | 50 mM Tris-HCl (pH=8.0), 1 M NaCl, 5 mM imidazole, 1 mM 2-mercaptoethanol, supplemented with EDTA-free Protease inhibitor cocktail (Roche) and 1 mM freshly prepared PMSF |
| His-AC washing buffer | 50 mM Tris-HCl (pH=8.0), 1 M NaCl, 50 mM imidazole, 1 mM 2-mercaptoethanol |
| His-AC elution buffer | 50 mM Tris-HCl (pH=8.0), 100 mM NaCl, 500 mM imidazole, 1mM 2-mercaptoethanol |
| SEC buffer | 50 mM Tris-HCl (pH=7.5), 100 mM NaCl, 10 mM MgCl2, 1 mM 2-mercaptoethanol |

**Table S5.** Parameters for the Shifted Smoothly-truncated Lennard-Jones model

| **Bead type** | **Energy parameter,** $\epsilon$ **(kcal mol$^{-1}$)** | **Distance parameter,** $\sigma\ (nm)$ |
|---|---|---|
| Small pole-like beads | 0.15 | 0.356 |
| Central excluded- volume beads | 0.15 | 2.60 |
| Larger neutral bead in 6-bead model | 0.15 | 6.00 |

**Movie S1 (separate file). Self-assembly video of 5-bead model for charge q = ±3.8 *e*.** Self-assembly of 10000 monomers forming filaments over the course of 12 µs in a 500x500x500 nm space. Largest filament size recorded has 15 monomers. Filaments that are less than half the max size recorded in data are made transparent and lighter in color with the rest darker in color and visible.

**Movie S2 (separate file). Self-assembly video of 5-bead model of charge q = ±4.0 *e*.** Self-assembly of 10000 monomers forming filaments over the course of 24 µs in a 500x500x500 nm space. Largest filament size recorded has 32 monomers. Filaments that are less than half the max size recorded in data are made transparent and lighter in color with the rest darker in color and visible.

**Movie S3 (separate file). Self-assembly video of 5-bead model of charge q = ±5.0 *e*.** Self-assembly of 10000 monomers forming filaments over the course of 24 µs in a 500x500x500 nm space. Largest filament size recorded has 82 monomers. Filaments that are less than half the max size recorded in data are made transparent and lighter in color with the rest darker in color and visible.

**SI References**